\documentclass[twocolumn]{aastex62}
\usepackage{multirow}
\usepackage{amsmath}
\usepackage{graphicx}

\graphicspath{{./}{figures/}}

\begin{document}

\title{Optimizing simulation parameters for weak lensing analyses involving non-Gaussian observables}

\correspondingauthor{Jos\'e Manuel Zorrilla Matilla}
\email{jzorrilla@astro.columbia.edu}

\author[0000-0002-6267-716X]{Jos\'e Manuel Zorrilla Matilla}
\affiliation{Department of Astronomy, Columbia University, New York, USA}

\author{Stefan Waterval}
\affiliation{Department of Physics, ETH, Z\"urich, Switzerland}
\affiliation{Department of Physics, New York University Abu Dhabi, Abu Dhabi, United Arab Emirates}

\author{Zolt\'an Haiman}
\affiliation{Department of Astronomy, Columbia University, New York, USA}

\begin{abstract}
We performed a series of numerical experiments to quantify the sensitivity of the predictions for weak lensing statistics obtained in raytracing DM-only simulations, to two hyper-parameters that influence the accuracy as well as the computational cost of the predictions: the thickness of the lens planes used to build past light-cones and the mass resolution of the underlying DM simulation. The statistics considered are the power spectrum and a series of non-Gaussian observables, including the one-point probability density function, lensing peaks, and Minkowski functionals. 
Counter-intuitively, we find that using thin lens planes ($< 60~h^{-1}$\,Mpc on a $240~h^{-1}$\,Mpc simulation box) suppresses the power spectrum over a broad range of scales beyond what would be acceptable for an LSST-type survey. A mass resolution of $7.2\times 10^{11}~h^{-1}$\,$M_{\odot}$ per DM particle (or 256$^3$ particles in a ($240~h^{-1}$\,Mpc)$^3$ box) is sufficient to extract information using the power spectrum and non-Gaussian statistics from weak lensing data at angular scales down to 1\,arcmin with LSST-like levels of shape noise.
\end{abstract}

\section{Introduction} \label{sec:intro}

Weak gravitational lensing (WL) enables the mapping of the distribution of dark matter (DM) in the universe on large scales and as a result is a powerful probe to infer cosmological parameters such as $\sigma_8$ and $\Omega_{\text{m}}$ (see comprehensive reviews by, e.g. \citealt{Bartelmann2001},  \citealt{HoekstraHenkJain2008} and \citealt{Kilbinger2015}). In practice, the lensing signal can be extracted from statistical measurements of the shapes of background galaxies, distorted by deflections in the path of light rays as they traverse the vicinity of matter over- and under-densities.

Upcoming surveys such as DESI \citep{DESI2016}, LSST \citep{LSST2009}, Euclid \citep{Euclid2010}, WFIRST-AFTA \citep{WFIRST2015} and SKA \citep{SKA2018}, will provide WL data of unprecedented quality and quantity and will require correspondingly accurate and precise models to extract information from these datasets. WL observables delve into the non-linear regime, which can be forward-modeled by ray-tracing photons through high-resolution simulated dark matter (DM) density fields \citep{Schneider1988, Jain2000, Vale2003, Hilbert2009, Heitmann2010}. Simulating the large volumes encompassed by future surveys is computationally expensive, especially when a high-dimensional parameter space needs to be explored. Different ideas to reduce the cost of forward modeling WL observables have been put forward and tested. These include using approximate codes to simulate the evolution of the matter density field -- for example ICE-COLA \citep{Izard2018}, L-PICOLA \citep{Howlett2015} or FastPM \citep{Feng2016}--, analytic or semi-analytic models -- for example \textsc{Camelus} \citep{Lin2015} and machine learning approaches -- for example generative adversarial networks (GANs; e.g. \citealt{He2018, Rodriguez2018,Mustafa2019}). While analytic models can predict two-point statistics with sufficient accuracy \citep{Barreira2018}, higher-order statistics which capture non-Gaussian information from the non-linear regime require a numerical approach \citep{Sato2009, Petri2013,Petri2017}.

In this paper we study, within the framework of raytracing N-body simulations, the sensitivity of the power spectrum (PS) and several non-Gaussian statistics commonly used in WL studies: the one-point probability density function (PDF), lensing peak counts, and the full set of Minkowski functionals (MFs) to the two hyper-parameters with the highest impact on the computational cost of the simulations: (i) the thickness of the lens planes used to construct the past light cones in raytracing and (ii) the mass resolution of the N-body simulations used to model the underlying matter density field. Previous studies have already tackled some of these or related aspects. For instance, \cite{Jain2000} verified the effect on the measured PS of the mass resolution of the underlying DM simulation, the grid size and interpolation method used when raytracing and also studied the contribution of super-sample modes to the PS variance. \cite{Sato2009} looked at the effect of the resolution of the 2D lens planes on the measured convergence power spectrum, and evaluated the non-Gaussian contribution to its covariance matrix. \cite{Vale2003} investigated the effect of mass resolution and comoving distance between lens planes on the convergence power spectrum, skewness and kurtosis. This work revisits the sensitivity of the measured convergence power spectrum to the mass resolution and distance between lens planes, and extends the analysis of numerical convergence analyses to non-Gaussian statistics within the framework of a tomographic WL analysis of a LSST-like survey.

The manuscript is organized as follows. In \S~\ref{sec:methods} we describe our simulation pipeline (\S~\ref{subsec:model}), how we quantify the impact of the hyper-parameters (\S~\ref{subsec:method}), and the statistics used to assess that impact (\S~\ref{subsec:observables}). In \S~\ref{sec:results} we report and discuss the results of the analysis. Finally, we summarize our conclusions in \S~\ref{sec:conclusions}.

\section{Methods}\label{sec:methods}

\subsection{Simulating convergence maps} \label{subsec:model}

Our analysis is based on lensing statistics measured on convergence maps obtained from ray-tracing dark matter only N-body simulations. Since weak lensing probes large-scale structures in the non-linear regime, direct simulations offer a way to characterize the WL signal.

The N-body simulations are run using the publicly available Tree-PM code {\tt Gadget2} \citep{Springel2005},  which evolves Gaussian initial conditions generated with {\tt NGenIC} \citep{Springel2015}. The initial conditions are defined by power spectra computed with {\tt CAMB} \citep{Lewis2000}. The positions of the particles at different redshifts are used to build past light-cones. The trajectory of a bundle of rays is followed along past light-cones according to the multi-plane algorithm to generate synthetic convergence maps. We refer the reader to \cite{Petri2016} for a detailed step-by-step description of our pipeline and its implementation in {\tt LensTools},  and to \cite{Jain2000} for a review of the theoretical basis of the algorithms used. 

All the simulations have the same comoving volume of ($240h^{-1}$\,Mpc)$^3$, phases for the initial conditions, and underlying cosmology. The cosmological parameters are consistent with \cite{Planck2015}: ($H_0$, $\Omega_{\text{m}}$, $\Omega_{\Lambda}$, $\Omega_{\text{b}}$, $w_0$, $\sigma_8$, $n_{\text{s}}$) = (67.7\,km~s$^{-1}$Mpc$^{-1}$, 0.309, 0.691, 0.0486, -1.0, 0.816, 0.967). As an illustration, we compare the convergence power spectrum measured in our highest-resolution simulation ($1024^3$ DM-only particles, or a particle mass of $1.1\times10^{10}\,{\rm M_{\odot}}$), with flat-sky approximation predictions (see Eq.~\eqref{eq:ps}) computed using the {\tt halofit} model~\citep{Smith2003, Takahashi2012}, as implemented in the
Core Cosmology Library (CCL;~\citealt{CCL2019}). In data smoothed at $1\,{\rm arcmin}$ resolution with the presence of the shape noise levels considered here, the agreement between theory and simulations stays within 20\% for $\ell < 12,000$. In noiseless data, a loss of power in the simulations is noticeable, especially for sources at low redshift. The worst case corresponds to the bin with sources at $z=0.5$, which shows a 50\% loss in power at $\ell\approx7,500$. In all cases we measure excess power at large angular scales relative to the theoretical expectation (up to $\approx20\%$ for the noiseless case and $\approx13\%$ in the presence of shape noise). The reason of this excess is the sample variance in the matter power spectrum, as low multipoles probe scales that are comparable with the size of our simulated volume. Since the purpose of this study is the calibration of hyper-parameters, we are interested in relative changes and the sample variance at large scales should not impact our conclusions. Furthermore, we are mostly interested in relative changes on small scales, where the use of non-Gaussian observables is relevant. To perform inference from data, this sample variance would need to be tackled, for instance, by simulating a larger volume (or many independent N-body realizations of our simulation volume).

\begin{figure}
\begin{center}
\includegraphics[width=\linewidth]{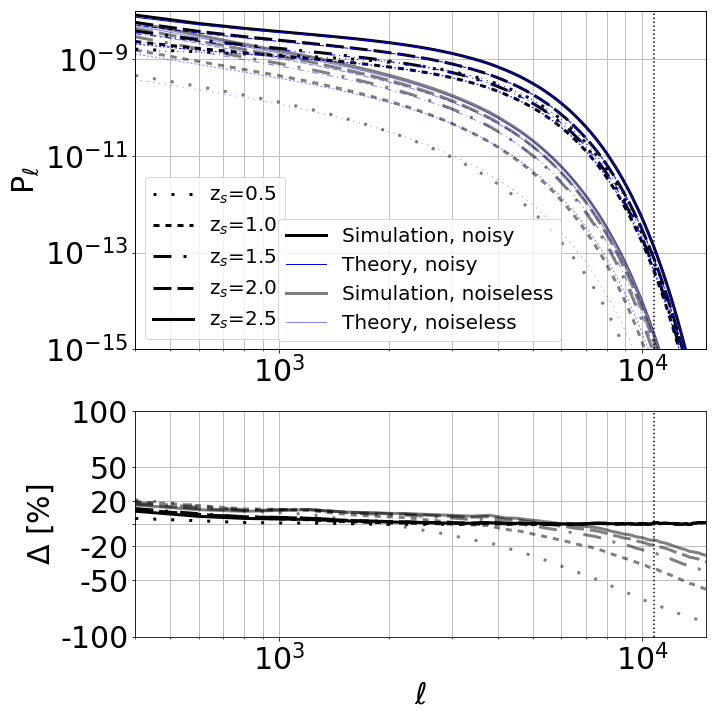}
\caption{Comparison of convergence power spectra measured in our simulations with $1024^3$ DM particles, to those obtained in semi-analytic calculations using the
flat-sky approximation and {\tt halofit}.
\textbf{Upper panel:} Convergence power spectra. Thick, black lines correspond to simulations, while thin, blue lines correspond to the {\tt halofit} calculations. Different dashes indicate different redshift bins for the lensed galaxies. Full lines correspond to results in the presence of shape noise (see section ~\S~\ref{subsec:model} for a description of the shape noise level considered), and partly transparent lines to the results in the absence of noise.
\textbf{Lower panel:} percentage difference between  simulations and {\tt halofit}, for each redshift bin. As in the upper panel, full lines correspond to the noisy, and partly transparent lines to the idealized noiseless case.
}
\label{fig:resolution}
\end{center}
\end{figure}

For each configuration, a single volume is simulated and reused through random shifts and rotations to generate as many as $\mathcal{O}\left(10^4\right)$ pseudo-independent past light-cones \citep{Petri2016Variance}. Since each convergence map covers only $3.5\times3.5$\,deg$^2$ on the sky at this redshift, the flat-sky approximation holds.  To account for their intrinsic ellipticity, Gaussian random shape noise is added independently at each of the $1024\times1024$ pixels in each convergence map. The standard deviation of this noise,
\begin{equation}\label{eq:shape_noise}
\sigma_{\text{pix}} = \sqrt{\frac{\sigma_{\varepsilon}^2}{2n_{\text{g}}A_{\text{pix}}}},
\end{equation}
depends on the variance of the galaxies' intrinsic ellipticity (assumed to be $\sigma_{\epsilon}=0.35$), the solid area covered by a pixel ($A_{\text{pix}}=0.04$\,arcmin$^2$), and the effective galaxy number density. 

We consider an LSST-like survey with a footprint of $20,000\,{\rm deg^2}$ and  galaxy density distribution $n(z)\propto z^2 \exp \left[-2z\right]$~\citep{LSST2009}. Because a small number of redshift bins suffices to extract most of the tomographic information encoded in weak lensing data sets~\citep{Hu1999}, we use five tomographic bins. In each bin, we assume all lensed galaxies are at a fixed redshift of $\left[0.5, 1.0, 1.5, 2.0, 2.5\right]$. The galaxy densities we consider in each redshift bin are $\left[8.83, 13.25, 11.15, 7.36, 4.26\right]\,{\rm arcmin}^{-2}$.

A final smoothing is applied to both noiseless and noisy maps with a Gaussian kernel of 1\,arcmin standard deviation. While only results from maps with shape noise are relevant for the analysis of future survey data, we show also results for smoothed, noiseless convergence maps, since the effect of different simulation choices are often more discernible in those.

\subsection{Assessing the impact of hyper-parameters} \label{subsec:method}

The accuracy of the forward model as a function of different values of the hyper-parameters is assessed by comparing the statistics of observables measured over 10,048 convergence maps simulated for each configuration. As explained in \S~\ref{subsec:model}, for each set of hyper-parameters, all 10,048 maps are generated from a single, recycled, N-body simulation.

We chose a "fiducial" configuration as a reference. The difference between an observable's mean for all cases and the fiducial model is compared with a standard error. As standard error, we consider 3 standard deviations measured on the fiducial model's maps, scaled to a survey sky area of $2\times10^4$\,deg$^2$ (commensurate with LSST's). This scaled standard deviation represents a lower bound on the uncertainty expected in future surveys, as it includes only the statistical error from intrinsic ellipticities. For a review of all sources of error in WL surveys, see for instance \cite{Shirasaki2014}.

Ultimately, the most relevant metric is how much the lens plane thickness affects the inference of cosmological parameters. A definite answer to this question requires the calculation of the credible contours for the parameters of interest. Within the scope of our single-cosmology numerical experiments, we instead look at the reduced $\chi^2$, which combines differences in the mean observables with their covariance matrix.

In general, a given observable is a vector $\mathbf{s}$ with components corresponding to bins of spherical harmonic index $\ell$ for the power spectrum, or $\kappa$ thresholds for the other statistics, such as the one-point PDF, lensing peaks or Minkowski functionals. We compute the $\chi^2$ statistic for each configuration, considering the mean of the fiducial model as ground truth:
\begin{equation}\label{eq:chi2}
    \chi^2 = \left(\mathbf{s}-\mathbf{s}_{fid}\right)^{T}\widehat{\mathsf{C}^{-1}}\left(\mathbf{s}-\mathbf{s}_{fid}\right),
\end{equation}
where $\widehat{\mathsf{C}^{-1}}$ is an unbiased estimator of the precision matrix. We compute (and report) results based both on the precision matrix estimated at the fiducial model, and the precision matrix estimated at each specific configuration. We used the prescription from \cite{Hartlap2007} to de-bias the estimator of the precision matrix. For each observable, we report the $\chi^2$ per degree of freedom. Furthermore, to account for the effect of a change in configuration on the covariance matrix, we report two values of the reduced $\chi^2$, one based on the covariance matrix for the reference configuration and another based on the configuration-specific covariance. We consider two configurations as statistically equivalent when their $\chi^2$ per degree of freedom, in the presence of noise, is less than or equal to unity.

\subsection{Hyper-parameter configurations}\label{subsec:configurations}

We analyzed different configurations for two hyper-parameters: the thickness of the lensing planes used in the ray-tracing algorithm, and the mass resolution (number of particles) of the N-body simulations. For each configuration, we ray-traced 10,048 convergence maps, each of these maps with an area of $12.25\,{\rm deg}^2$.

The positions of the DM particles were saved at redshifts that allowed the construction of light-cones with lens planes at comoving distances of $20~h^{-1}$, $40~h^{-1}$, $60~h^{-1}$ $80~h^{-1}$, and $120~h^{-1}$\,Mpc. Thinner lens planes can potentially capture more accurately the evolution of the matter density field with redshift. However the number of planes needed to cover a redshift range increases as the plane thickness decreases, and so do the computational and storage requirements for the simulations. In particular, the two tasks that account for the largest increase in computation time are the calculation of the gravitational potential at the planes (solving a 2D Poisson equation in Fourier space) and the computation of the Jacobian matrix that determines the light ray's deflections at each point on the planes \citep{Petri2016}. The fiducial case corresponds to a lens plane thickness of $80~h^{-1}$\,Mpc, a value that has been typically used in prior work \citep{Yang2011, Petri2013, Jose2016}, provides 9 independent lens planes per simulation snapshot (increasing the number of pseudo-independent $\kappa$ maps that can be generated from a single N-body simulation), and is not large enough to show discreteness effects with lensed galaxies at $z=2$ \citep{Jain2000}.

The minimum angular scale at which cosmological information can be extracted is limited by the depth of the survey, which determines the number density of galaxies whose shape can be measured, and the accuracy of the forward models used to predict the signal. Baryonic physics~\citep{Huang2018} and intrinsic alignments~\citep{Chisari+2015} restrict the accuracy of current models at small scales. Matching the mass resolution of the underlying N-body simulations to the scales at which the analysis of the data is reliable saves computational resources.

In our fiducial N-body simulations we used $1024^3$ particles, and we ran additional simulations with $128^3$, $256^3$ and $512^3$ particles, which yield mass resolutions per DM particle of $5.7\times 10^{12}\,h^{-1}{\rm M_{\odot}}$, $7.2\times 10^{11}\,h^{-1} {\rm M_{\odot}}$, $9.0\times 10^{10}\,h^{-1}{ \rm M_{\odot}}$ and $1.1\times 10^{10}\,h^{-1}{\rm M_{\odot}}$. We note that due to memory limitations, we increased the number of CPUs allocated to our higher-resolution simulation, resulting  in a worse scaling than the one that can be achieved in an optimized setup \citep{Springel2015}.

Tables \ref{table:cputime} and \ref{table:filesize} summarize the computational cost of the main steps involved in our simulation pipeline, and the disk space required for storing the different data products (in practice, not all need to be saved). Performance benchmarks are based on runs using Intel Knights Landing nodes from TACC's Stampede2 supercomputer at the NSF XSEDE facility.

\begin{table*}
\begin{center}
\begin{tabular}{ccccc}
\hline
\hline
\multicolumn{1}{c}{} & \multicolumn{1}{c}{Task} & \multicolumn{1}{c}{RAM} &  \multicolumn{1}{c}{CPU Time} & \multicolumn{1}{c}{Change CPU [\%]}\\
\hline
\multicolumn{1}{c}{Plane thickness} & & &\\
\hline
\multicolumn{1}{l}{20 $h^{-1}$Mpc} & \multirow{5}{*}{lens planes} & \multirow{5}{*}{96 GB} & 82.7 & +1450\%\\
\multicolumn{1}{l}{40 $h^{-1}$Mpc} &  & & 21.3 & +300\%\\
\multicolumn{1}{l}{60 $h^{-1}$Mpc} &  & & 9.3 & +75\%\\
\multicolumn{1}{l}{$\mathbf{80 h^{-1}}$ \textbf{Mpc}} &  & & \textbf{5.3}\\
\multicolumn{1}{l}{120 $h^{-1}$Mpc} & &  & 2.7 & -50\%\\
\hline
\multicolumn{1}{l}{20 $h^{-1}$Mpc} & \multirow{5}{*}{Ray tracing} & \multirow{5}{*}{384 GB} & 1621.3 & +407\%\\
\multicolumn{1}{l}{40 $h^{-1}$Mpc} &  &  & 640 & +100\%\\
\multicolumn{1}{l}{60 $h^{-1}$Mpc} &  &  & 426.7 & +33\%\\
\multicolumn{1}{l}{$\mathbf{80 h^{-1}}$ \textbf{Mpc}} &  &  & \textbf{320}\\
\multicolumn{1}{l}{120 $h^{-1}$Mpc} &  &  & 213.3 & -33\%\\
\hline
\multicolumn{1}{c}{\# of particles} & & &\\
\hline
\multicolumn{1}{l}{$128^3$} & \multirow{5}{*}{N-body} & 576\,GB & 736 & -99\%\\
\multicolumn{1}{l}{$256^3$} &  & 576\,GB & 1408 & -98\%\\
\multicolumn{1}{l}{$512^3$} &  & 576\,GB & 6144 & -94\%\\
\multicolumn{1}{l}{$\mathbf{1024^3}$} &  & \textbf{9216\,GB} & \textbf{103833} \\
\hline
\end{tabular}
\caption{Memory and computational time requirements for main simulation tasks for different hyper-parameter values. Each CPU time unit is a core hour (representing wall clock time if computed in series). Changes in CPU time are relative to the fiducial run (in bold).}
\label{table:cputime}
\end{center}
\end{table*}

\begin{table*}
\begin{center}
\begin{tabular}{ccccccc}
\hline
\hline
\multicolumn{1}{c}{} & \multicolumn{1}{c}{Snapshot size} & \multicolumn{1}{c}{\# of snapshots} & \multicolumn{1}{c}{Total memory} &  \multicolumn{1}{c}{Plane size} & \multicolumn{1}{c}{\# of planes} & \multicolumn{1}{c}{Total memory}\\
\hline
\multicolumn{1}{c}{Plane thickness} & & & & & &\\
\hline
\multicolumn{1}{l}{20 $h^{-1}$Mpc} & \multirow{5}{*}{28.8 GB} & 226 & 792\,GB & \multirow{5}{*}{65\,M} & 8064 & 524.2\,GB\\
\multicolumn{1}{l}{40 $h^{-1}$Mpc} &  & 114 & 400\,GB &  & 2016 & 131\,GB\\
\multicolumn{1}{l}{60 $h^{-1}$Mpc} &  & 77 & 270\,GB &  & 900 & 58.5\,GB\\
\multicolumn{1}{l}{$\mathbf{80 h^{-1}}$\,\textbf{Mpc}} &  & \textbf{58} & \textbf{204\,GB} &  & \textbf{504} & \textbf{32.8\,GB}\\
\multicolumn{1}{l}{120 $h^{-1}$Mpc} &  & 40 & 141\,GB &  & 228 & 14.8\,GB\\
\hline
\multicolumn{1}{c}{\# of particles} & & & & &\\
\hline
\multicolumn{1}{l}{$128^3$} & 55\,M & \multirow{4}{*}{58} & 3.2\,GB & \multirow{4}{*}{65\,M} & \multirow{4}{*}{504} & \multirow{4}{*}{32.8\,GB}\\
\multicolumn{1}{l}{$256^3$} & 448\,M &  & 26\,GB &  & &\\
\multicolumn{1}{l}{$512^3$} & 3.5\,GB &  & 204\,GB & & &\\
\multicolumn{1}{l}{$\mathbf{1024^3}$} & \textbf{28.8\,GB} & & \textbf{1670.4\,GB} & & &\\
\hline
\end{tabular}
\caption{Individual plane and snapshot size and the respective total storage requirements for both intermediate data products. In bold, values for the fiducial configuration.}
\label{table:filesize}
\end{center}
\end{table*}

\subsection{Observables}\label{subsec:observables}

We analyze the effect of changes in the hyper-parameters used to generate the synthetic data on the convergence power spectrum and a series of non-Gaussian observables: the one-point probability density function (PDF), lensing peak counts, and the full set of Minkowski functionals (MFs).

The power spectrum is the Fourier transform of the two-point correlation function, encodes all the information available in a Gaussian random field, and can be accurately predicted from theory on large angular scales. Therefore, it is commonly used in WL analyses \citep{DES2005, HSC2018}. We measured it in 20 logarithmic bins between $\ell_{min}=200$ and $\ell_{max}=12,000$, extending slightly above the smoothing scale of $1\,{\rm arcmin}$ which corresponds to $\ell \approx 10,800$. The relatively featureless power spectra can be properly characterized with 20 bins, without yielding a very large data vector when combining the 5 auto-spectra with the 15 cross-spectra for all 5 redshift bins. Since the synthetic maps cover a small field of view, we can use the flat-sky approximation to model the expected power spectrum from measuring the ellipticities of lensed sources at a fixed radial comoving distance, $\chi_s$:

\begin{equation}\label{eq:ps}
    P_{\kappa} \left(\ell \right) = \left(\frac{3H_0^2}{2c^2}\Omega_m\right)^2 \int_0^{\chi_s}\frac{d\chi}{a^2\left(\chi\right)}\left(1-\frac{\chi}{\chi_s}\right)^2 P_{\delta}\left(k=\frac{\ell}{\chi}; \chi\right).
\end{equation}

Here $P_{\delta}$ is the (non-linear) matter power spectrum, which depends on the wavenumber $k$ and evolves with redshift, and $H_0$, $\Omega_m$ and $c$ are the Hubble constant, matter density parameter and speed of light, respectively. Shape noise contributes white noise with total power of $P_{\epsilon}=\sigma_{\epsilon}^2/n_g$. Smoothing with a Gaussian kernel of width $\sigma_s$ multiplies the convergence power spectrum by the beam function $b_{\ell}=\exp{\left[-\ell\left(\ell+1\right)\sigma_s^2\right]}$.

The effect that the finite thickness of the lens planes has on the measured convergence power spectrum can be modelled as convolving the matter density power spectrum with a window function. Following~\citep{Takahashi2017}, it suffices to substitute  $P_{\delta}$ in Eq.~\eqref{eq:ps} by
\begin{equation}\label{eq:ps_thickness}
    P'_{\delta}\left(k; \chi\right) = \frac{\Delta \chi}{\pi} \int_0^{k^{\rm max}_{\parallel}} dk_{\parallel}P_{\delta}\left(k\right){\rm sinc}^2\left(\frac{k_{\parallel}\Delta \chi}{2}\right),
\end{equation}
where $k=\sqrt{k_{\parallel}^2+k_{\perp}^2}$, $k_{\perp}=\ell/\chi$, ${\rm sinc}(x)=\sin x/x$, and $\Delta \chi$ is the thickness of the lens planes used to build past light-cones in our simulations.

Changes in the mass resolution of the N-body simulations will also affect the measured $\kappa$ power spectrum. On one hand, the matter power will be suppressed on scales below the simulation resolution. On the other hand, an additional shot-noise component will result from the discreteness of the simulated matter density field. Both effects can be accounted for as in~\citep{Vale2003}, by smoothing the matter power spectrum with a Gaussian kernel of with $\sigma_N$ and adding a Poisson noise contribution corresponding to the finite number density $n_p$ of simulated particles,

\begin{equation}\label{eq:ps_particles}
    P''_{\delta}\left(k;\chi\right) = P_{\delta}\left(k;\chi\right)\exp{\left(-\sigma_N^2k^2\right)+\frac{1}{n_p}}.
\end{equation}

Most scales in our simulated maps probe non-linear structures, and as a result, there is non-Gaussian information not captured by the convergence power spectrum. To assess the impact of the different simulation parameters on the non-Gaussian information content of convergence maps, we analyze the effect on the one-point probability density function, the distribution of lensing peaks, and the Minkowski functionals.

The convergence PDF has been shown to help break the power spectrum degeneracy in the $\Omega_m-\sigma_8$ plane, improving cosmological constraints from the power spectrum alone~(e.g.~\citealt{Wang+2009,Patton2017}). It has also been used in order to assess the ability of weak lensing measurements to infer the neutrino masses~\citep{Liu2019}. We use the binned PDF measured on our (smoothed) simulated maps within the range $\left[-3\sigma_{\kappa}, 5\sigma_{\kappa}\right]$, where $\sigma_{\kappa}$ is the standard deviation of the maps corresponding to the highest mass resolution ($1024^3$ particles), and the fiducial, $80\,h^{-1}\,{\rm Mpc}$, lens planes. For each redshift bin in the noiseless case, we use $\sigma_{\kappa} = \left[0.007, 0.014, 0.019, 0.023, 0.027 \right]$, and for the maps that include shape noise $\sigma_{\kappa} = \left[0.034, 0.031, 0.035, 0.043, 0.055\right]$. We compare the measurements over simulations with the analytic expectation for Gaussian random fields (GRFs) with variance
\begin{equation}
    \sigma_0^2=\int_0^{\infty}\frac{\ell d\ell}{2\pi}P_{\kappa}\left(\ell\right).
\end{equation}\label{eq:pdf}
Lensing peaks are local maxima of the convergence field; their distribution as a function of their height, $\kappa$, was proposed to constrain cosmological parameters~\citep{Kratochvil2010, Dietrich2010}, and has been used extensively and successfully ever since~\citep{Liu2015, LiuPan2015, Kacprzak2016, Martinet2018}. We measured peak histograms in 50 linear bins covering the range $\kappa \in \left[-0.5, 5.0\right]$ in units of the r.m.s. shape noise in each redshift bin. For the maps smoothed on $1\,{\rm arcmin}$ scales, the r.m.s. of the shape noise for the redshift bins $\left[0.5, 1.0, 1.5, 2.0, 2.5\right]$ is $\sigma_{\rm n}=\left[0.033, 0.027, 0.030, 0.036, 0.048 \right]$. As we did for the PDF, we compare measurements with the predictions for GRFs with the same convergence power spectrum as the simulated $\kappa$ maps. The analytic expressions for the number of peaks in GRFs was derived in \cite{Bond1987}. We reproduce here the relevant equations for convenience. The number of lensing peaks in a GRF for a given height $\nu=\kappa/\sigma_0$ is given by:
\begin{equation}
    n_{\rm peaks}\left(\nu\right) = \frac{1}{\left(2\pi\right)^{3/2}\theta^2}\exp{\left(-\frac{\nu^2}{2}\right)}G\left(\gamma, \gamma \nu \right),
\end{equation}
where $\theta=\sqrt{2}\sigma_1/\sigma_2$ and $\gamma=\sigma_1^2/(\sigma_0\sigma_2)$ depend on the moments of the convergence power spectrum
\begin{equation}
    \sigma_p^2\int_0^{\infty}\frac{\ell d \ell}{2\pi}\ell^{2p}P_{\kappa}\left(\ell\right)
\end{equation}
and
\begin{multline}
     G\left(\gamma, x\right) = \left(x^2-\gamma^2\right)\left[1-\frac{1}{2}{\rm erfc}\left(\frac{x}{\sqrt{2\left(1-\gamma^2\right)}}\right)\right]  + \\
     x\left(1-\gamma^2\right)\frac{\exp{\left(-\frac{x^2}{2\left(1-\gamma^2\right)}\right)}}{\sqrt{2\pi\left(1-\gamma^2\right)}} + \\
     \frac{\exp{\left(-\frac{x^2}{3-2\gamma^2}\right)}}{\sqrt{3-2\gamma^2}}\left[1-\frac{1}{2}{\rm erfc}\left(\frac{x}{\sqrt{2\left(1-\gamma^2\right)\left(3-2\gamma^2\right)}}\right)\right].
\end{multline}

Finally, the three Minkowski functionals that can be defined in a two-dimensional random field, $\left\{V_0,\,V_1,\,V_2\right\}$, measure the area, length, and genus of the subset of points in the random field whose value exceeds a given threshold~\citep{Mecke1994, Schmalzing1996}. While $V_0$ conveys the same information as the PDF, we include it in our measurements for completeness. They have been extensively used to measure or extract non-Gaussian information from cosmological datasets, including weak lensing~\citep{Munshi2012, Petri2013, Shirasaki2014, Ling2015, Marques2019}.

We measured the three MFs in 28 linear bins covering the range $\kappa \in \left[-2, 5\right]$ in units of the r.m.s. shape noise for each redshift bin, and compared it with analytical predictions for GRFs. The MFs of two-dimensional GRFs are determined by the moments of their power spectrum, and their expression as a function of $\nu$ can be found in~\cite{Matsubara2000}:

\begin{eqnarray}
    V_0\left(\nu\right) &=& \frac{1}{2}{\rm erfc}\left(\frac{\nu}{\sqrt{2}}\right) \\
    V_1\left(\nu\right) &=& \frac{1}{8\sqrt{2}}\frac{\sigma_1}{\sigma_0}\exp{\left(-\frac{\nu^2}{2}\right)} \\
    V_2\left(\nu\right) &=& \frac{1}{\left(2\pi\right)^{3/2}}\left(\frac{\sigma_1}{\sqrt{2}\sigma_0}\right)^2\nu\exp{\left(-\frac{\nu^2}{2}\right).}
\end{eqnarray}

While a perturbative expansion has been found for MFs for non-Gaussian random fields~\citep{Munshi2012}, they have been shown to converge too slowly at the relevant $\sim$arcmin smoothing scales~\citep{Petri2013}, so we do not investigate them here.

\section{Results and discussion} \label{sec:results}

We discuss the effect on lensing statistics of the comoving distance between lens planes (plane thickness) in  \S~\ref{subsec:thickness} and mass resolution in \S~\ref{subsecsec:resolution}.

\subsection{Lens plane thickness} \label{subsec:thickness}

\begin{figure*}
\begin{center}
\includegraphics[width=\linewidth]{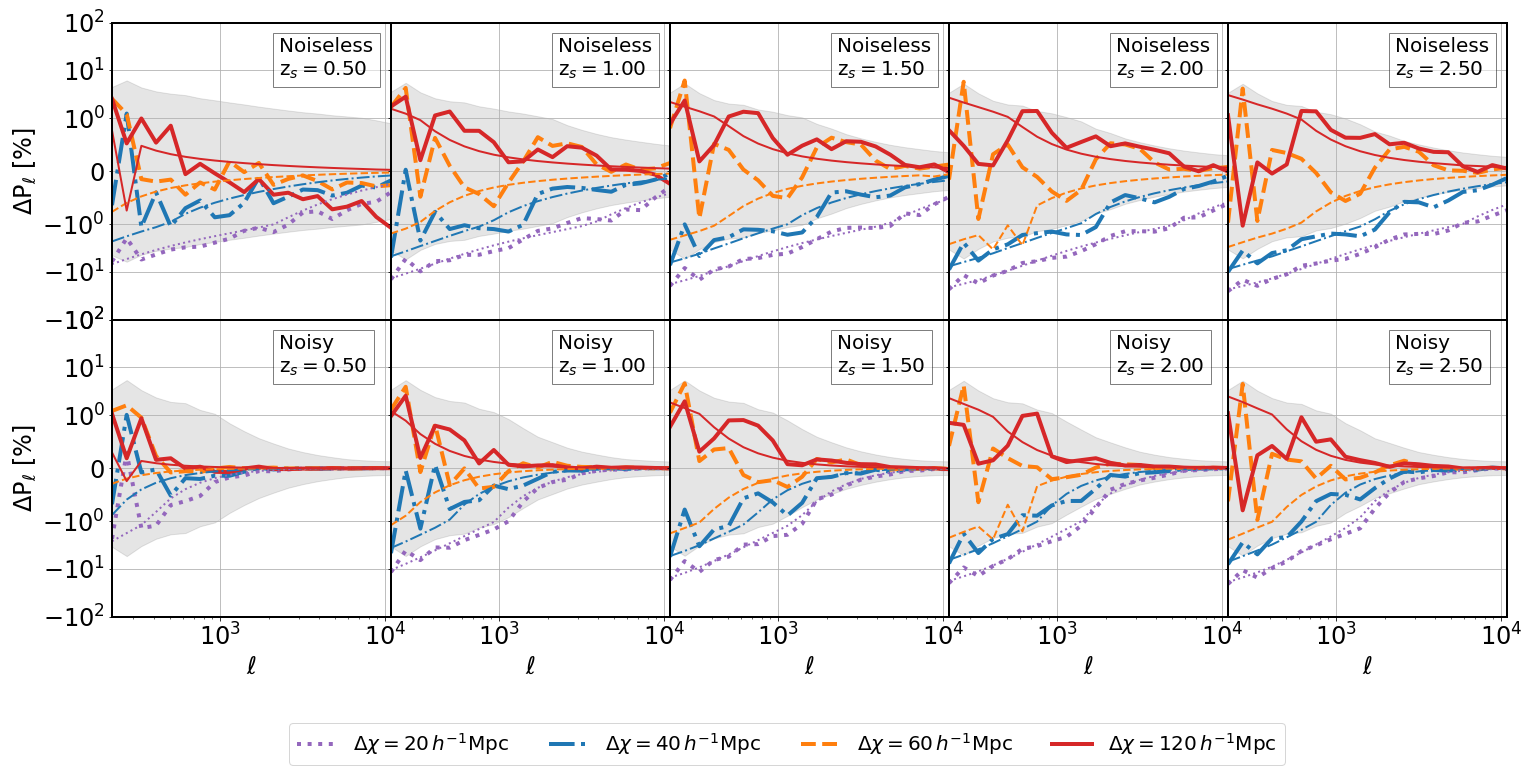}
\caption{Effect of lens plane thickness on convergence power spectra. Each column shows, for a different redshift bin, the percentage difference in the mean auto power spectrum measured over 10,048 convergence maps for a given thickness of the lens planes, relative to the fiducial value of $80~h^{-1}\,{\rm Mpc}$. 
The top row corresponds to noiseless data, and the bottom row to data in the presence of shape noise (in all cases, maps were smoothed at $1\,{\rm arcmin}$ resolution). Thick lines represent measurements over simulated maps, while thin lines represent theoretical predictions following~\cite{Takahashi2017}. For comparison, a standard error is shown in shaded gray, corresponding to 3 standard deviations of the measurements in the fiducial case, scaled to a $2\times10^4$\,deg$^2$ survey.}
\label{fig:planethickness_ps}
\end{center}
\end{figure*}

We show the mean percentage difference in the power spectrum for noiseless and noisy convergence maps for different lens plane thicknesses (or comoving distance between lens planes) relative to the fiducial case of $80~h^{-1}$\,Mpc in Figure~\ref{fig:planethickness_ps}. For each plane thickness, the power spectrum is measured and averaged over 10,048 convergence maps. For clarity, we do not display the results for the 10 cross-spectra between redshift bins, but the conclusions are not altered. The gray band in each panel represents a standard error corresponding to 3 times the standard deviation measured over the fiducial maps, scaled to a $2\times10^4$\,deg$^2$ LSST-like survey. This does not incorporate the effect of off-diagonal terms in the covariance matrix, which are fully utilized (including auto and cross spectra) in the $\chi^2$ statistic described in \S~\ref{subsec:method} to assess the significance of the differences between each configuration and the fiducial case.

The effect of the finite width of lens planes, measured on the simulations, matches well the expectation from the model described in ~\cite{Takahashi2017}, in which the matter power spectrum is convolved with the window function corresponding to a spherical lens shell. Since the differences between the fiducial and the $60~h^{-1}$\,Mpc case are in general sub-percent level, they fall in the linear range of the y-scale and appear to be larger than those for other cases, but they are not. As the lens planes cut structures, there is an appreciable loss of power for thin planes throughout the multipole range considered in this study. The presence of shape noise renders those differences statistically insignificant for the range of thicknesses $60~h^{-1} - 120~h^{-1}$\,Mpc (see Table~\ref{table:chi2_planes}).

\begin{figure*}
\begin{center}
\includegraphics[width=\linewidth]{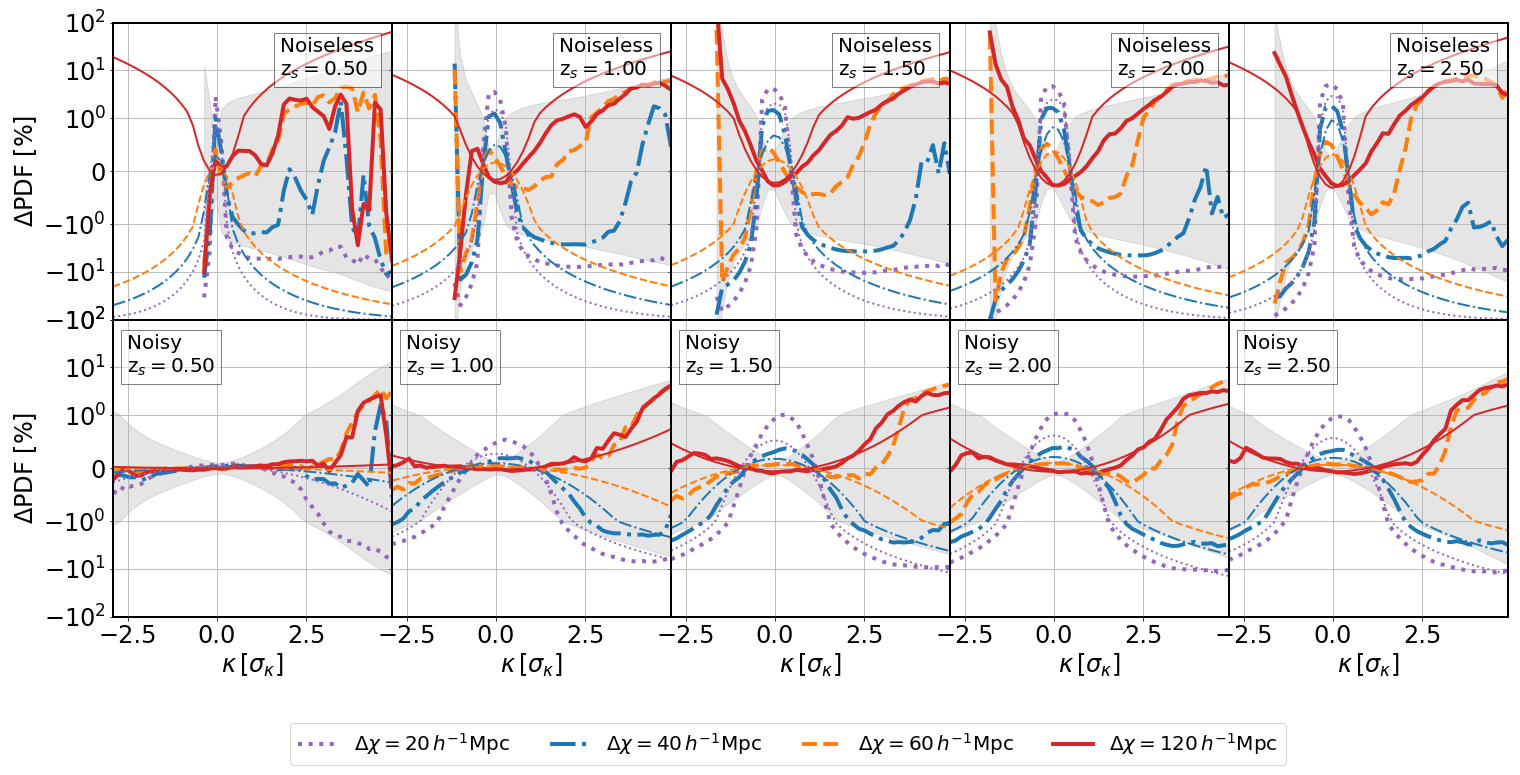}
\caption{Same as Figure~\ref{fig:planethickness_ps} for the one-point $\kappa$ probability density function (PDF). Thick lines correspond to measurements over simulated data, thin lines to predictions for Gaussian random fields with the same power spectrum as the simulated maps.}
\label{fig:planethickness_pdf}
\end{center}
\end{figure*}

The effect on the PDF is shown in Figure~\ref{fig:planethickness_pdf}, where the PDF change measured from maps built using different lens plane thicknesses is plotted (thick lines) together with predictions for Gaussian random fields (thin lines). The GRFs have the same power spectrum as the simulated $\kappa$ maps. The overall impact on the GRFs is straightforward: the loss in power translates into a smaller variance for the random field (see Eq.~\eqref{eq:pdf}). As a result we would expect, for planes thinner than the fiducial, a suppression in the tails of the PDF compensated by an enhancement near the peak of the PDF (and the opposite for a the case with thicker lens planes). Qualitatively, the measurements in the presence of shape noise follow that trend, which is to be expected since the shape noise model considered is Gaussian. In the limit of noiseless data, while there is still some qualitative agreement, the quantitative differences with the GRF model are evident.

From a practical perspective, lens planes whose thickness lays in the range of $60-120~h^{-1}$ Mpc yield convergence maps with statistically indistinguishable PDFs, even in the absence of shape noise (see Table~\ref{subsec:thickness}).

\begin{figure*}
\begin{center}
\includegraphics[width=\linewidth]{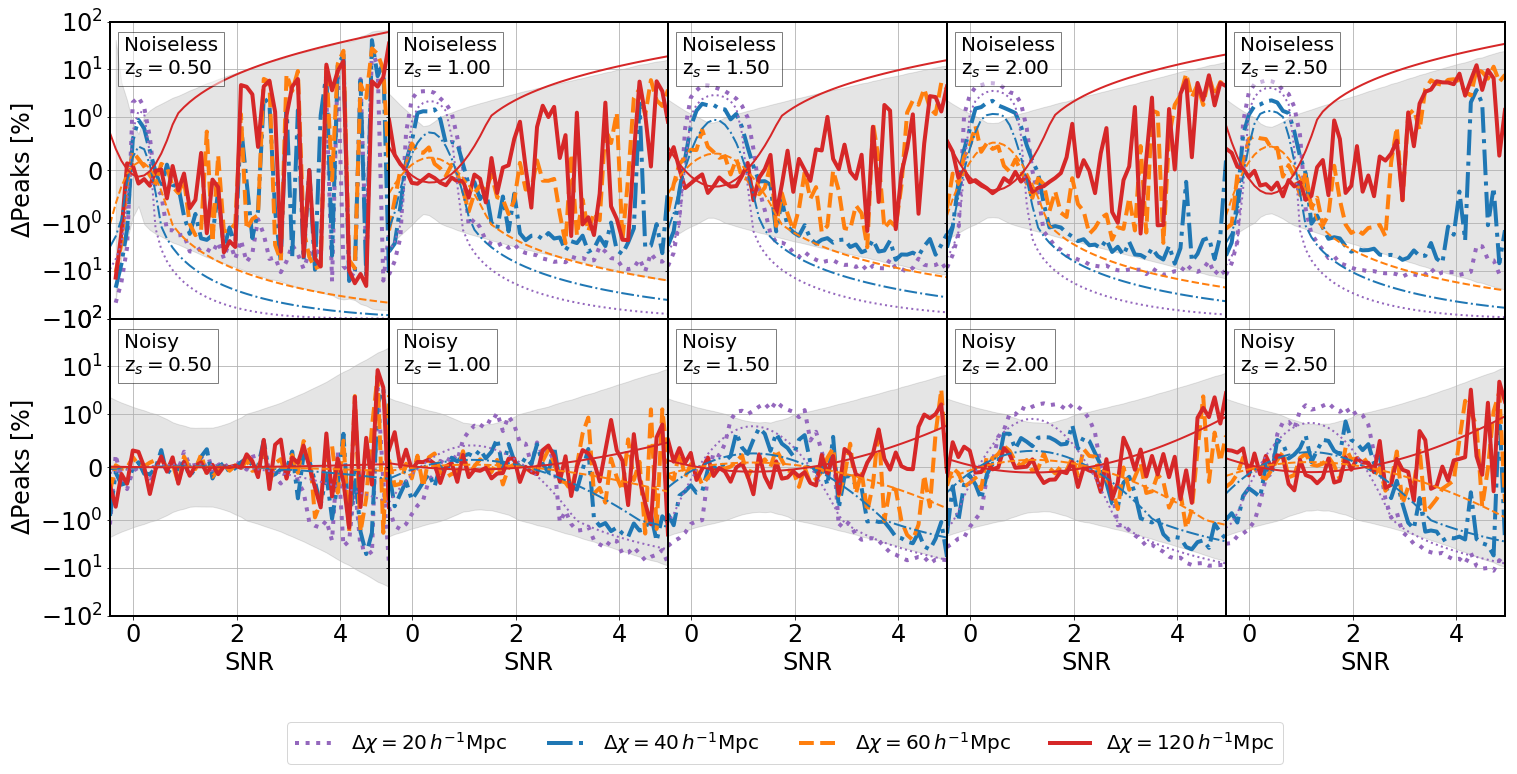}
\caption{Same as Figure~\ref{fig:planethickness_ps} for the lensing peak distribution. Thick lines correspond to measurements over simulated data, thin lines to predictions for Gaussian random fields with the same power spectrum as the simulated maps.}
\label{fig:planethickness_pc}
\end{center}
\end{figure*}

The sensitivity of the distribution of lensing peak counts to the width of the lens planes used during raytracing is shown in Figure~\ref{fig:planethickness_pc}. For noiseless data, lens planes thinner than $60~h^{-1}$ Mpc overproduce low-significance peaks while underproducing higher-significance ones. Table~\ref{table:chi2_planes} shows how the results from using lens planes within the range $60-120~h^{-1}$ Mpc are statistically indistinguishable. The addition of shape noise reduces the differences further, to the point where using thinner planes, at $40~h^{-1}$ Mpc, could be safe. As seen from the power spectrum, there is not a strong case to move from the fiducial thickness of $80~h^{-1}$ Mpc.

As for the PDF, the changes in the lensing peak counts in $\kappa$ maps are qualitatively similar to those in GRFs with the same power spectrum, but there are quantitative differences. The effect is also similar to the one seen for the PDF, and can be understood intuitively: thicker lens planes provide more power on larger scales that modulates the heights of local maxima, increasing the variance of their distribution (and conversely for thinner planes).

Past studies have found that the use of the Born approximation, in which the convergence is directly estimated weighting the projected matter density, can induce a significant bias to predictions of non-Gaussian observables such as the skewness and kurtosis~\citep{Petri2017}. Since in the limit of very thick planes, the multi-plane ray-tracing algorithm is similar to the Born approximation (except for the lack of redshift evolution in the matter density field within each plane), any attempt to predict peak histograms beyond the range of thickness explored in this study, would need to be validated.

\begin{figure*}
\begin{center}
\includegraphics[width=\linewidth]{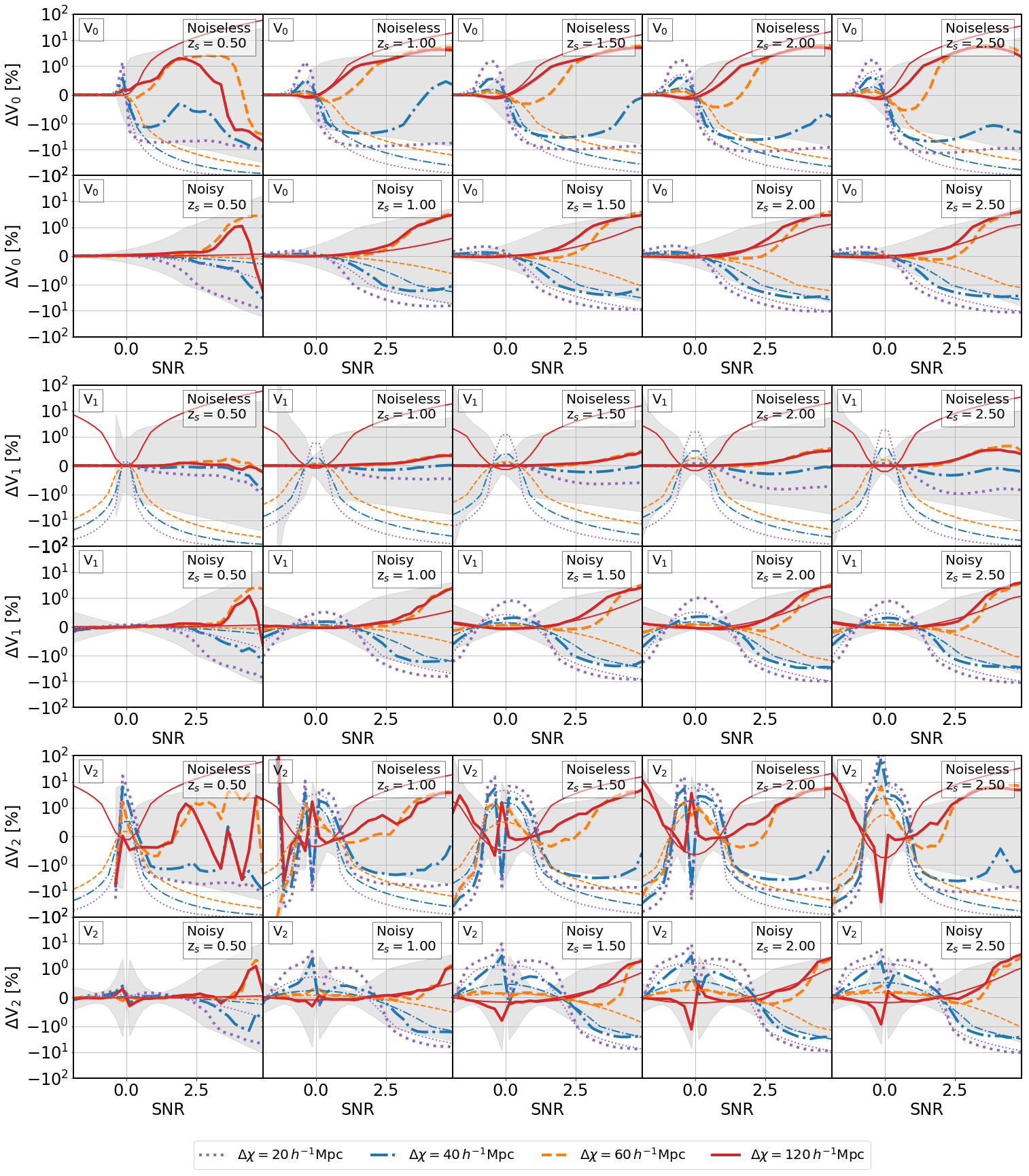}
\caption{Same as Figure~\ref{fig:planethickness_ps} for the three Minkowski functionals (MFs). Each pair of rows shows a different MF. The top row of each pair corresponds to noiseless data, and the bottom row to data in the presence of shape noise (in all cases, maps were smoothed at $1\,{\rm arcmin}$ resolution). Thick lines correspond to measurements over simulated data, thin lines to predictions for Gaussian random fields with the same power spectrum as the simulated maps.}
\label{fig:planethickness_mfs}
\end{center}
\end{figure*}

The effect of lens plane thickness on the MFs is different for each functional (see Figure~\ref{fig:planethickness_mfs}). As expected, shape noise "gaussianizes" the convergence field and as a result measured differences follow qualitatively the predictions for GRFs. This is not the case in the limit of noiseless data. The $\kappa$ field is less sensitive to changes in the plane thickness than what would be expected for GRFs, especially $V_1$. For $V_2$, the region of $\kappa\approx0$, where the functional changes sign, yields noisy measurements.

In terms of $\chi^2$ (Table~\ref{table:chi2_planes}), the MFs are more demanding than other non-Gaussian observables. While in the presence of shape noise, lens planes in the range of $60-120~h^{-1}$ Mpc yield measurements that are statistically equivalent, that is not the case anymore in the limit of noiseless data for $V_1$ and $V_2$.

\begin{figure}
\begin{center}
\includegraphics[width=\linewidth]{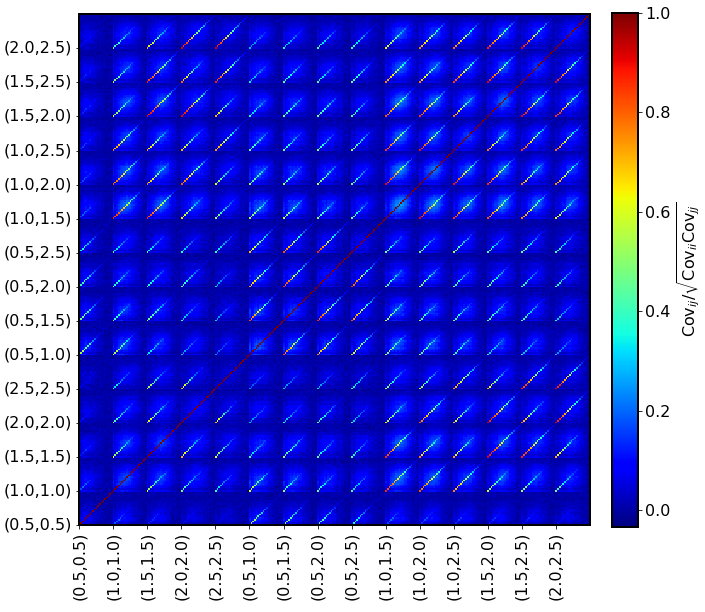}
\caption{Covariance matrix measured over 10048 convergence maps generated from our $512^3$ particle simulation with $80~h^{-1}\,{\rm Mpc}$ lens planes. The axis labels indicate the redshift bins used. For example, the slice [(1.0, 1.5),(2.5,2.5)] shows the correlation between the cross power spectrum of redshift bins $z_s=1.0$, $z_s=1.5$ and the auto power spectrum of the redshift bin $z_s=2.5$.}
\label{fig:planethickness_cov}
\end{center}
\end{figure}

As explained in \S~\ref{subsec:method}, the reduced $\chi^2$ allows us to assess the impact on inference, of the differences in the measured statistics from synthetic maps generated with different hyper-parameters. Changes in the covariance matrix are taken into consideration when computing $\chi^2$. We use either the covariance matrix from maps in the reference configuration, or an alternative covariance from maps recomputed after changing the thickness of the lens planes. In both cases, we use the full covariance to compute $\chi^2$, including, for instance, the 10 cross power spectra not shown in Figure~\ref{fig:planethickness_ps}. As an illustration, we display the full covariance matrix for the power spectrum measured on noisy maps in Figure~\ref{fig:planethickness_cov}.

The effect of lens planes in the covariance is small (for example, less than $20\%$ for the dominant, diagonal elements for the power spectrum). As a result, the conclusions are the same regardless of which covariance is used. Also, ~\cite{Barreira2018} have analyzed the impact of  non-Gaussian contributions to the lensing power spectrum covariance on parameter inference. They find that mean changes in the covariance matrix of $\approx 20\%$ translate to $\leq 5\%$ changes in the parameter uncertainties.

Table~\ref{table:chi2_planes} collects the results for the observables under study. With the shape noise levels considered, lens planes whose thickness is in the range of $60-120~h^{-1}$ Mpc yield statistically indistinguishable measurements for all the statistics we have studied, and are therefore safe to use. For a scheme, like ours, that recycles N-body snapshots to build different past light-cones, there is a trade-off between a loss in power induced by the lens plane's window function, and the number of lens planes that can be built from a single N-body simulation. The fiducial choice of $80~h^{-1}$\,Mpc seems reasonable, since it allows to generate as many as a few$\times 10^4$ pseudo-independent convergence maps from a single simulation~\citep{Petri2016Variance}. In the limit of noiseless data this is not the case anymore for the power spectrum and MFs. For the power spectrum, the effect of the lens plane thickness can be incorporated analytically.

These results do not take into consideration the uncertainty in the measured covariance matrices: if they are overestimated, our results could change. Nevertheless, the required errors must be substantial: the true covariance would need to be overestimated by more than $150\%$ (its true elements be smaller than $0.4\times$ the elements of the matrix measured on our simulations) for the Minkowski functionals (the most sensitive statistic to this hyper-parameter) measured on maps generated with $80~h^{-1}$\,Mpc lensing planes to be statistically distinguishable from those measured on maps generated with $120~h^{-1}$\,Mpc lensing planes.

\begin{table*}
\begin{center}
\begin{tabular}{lcccc}
\hline
\hline
& \multicolumn{2}{c}{Noiseless} & \multicolumn{2}{c}{Noisy}\\
\cline{2-5}
\multicolumn{1}{c}{Thickness} & \multicolumn{1}{c}{Model-dependent} & \multicolumn{1}{c}{Fiducial} & \multicolumn{1}{c}{Model-dependent} & \multicolumn{1}{c}{Fiducial}\\
\hline
& \multicolumn{4}{c}{Power spectrum reduced $\chi^2$}\\
\cline{2-5}
20 $h^{-1}$Mpc & 152.98 (188.31)& 98.78 (120.01) & 8.23 (11.72) & 5.59 (7.95)\\
40 $h^{-1}$Mpc &  17.33  (16.72) & 15.73  (14.74) & \textbf{0.83}  (1.17) & \textbf{0.74} (1.04) \\
60 $h^{-1}$Mpc &  6.96   (4.43) &  6.91   (4.34) & \textbf{0.17   (0.23)} & \textbf{0.17  (0.24)}\\
120 $h^{-1}$Mpc & 9.64   (5.25)  &  8.84   (5.09) &  \textbf{0.16   (0.21)} &  \textbf{0.16  (0.22)}\\
\cline{2-5}
& \multicolumn{4}{c}{PDF reduced $\chi^2$}\\
\cline{2-5}
20 $h^{-1}$Mpc  & 30.67 (40.27) & 18.02 (23.69) &  10.25  (13.31) & 7.50  (9.78)\\
40 $h^{-1}$Mpc  &  3.09  (3.96) &  2.51  (3.20) &  1.33  (1.66) & 1.23  (1.53)\\
60 $h^{-1}$Mpc  &  \textbf{0.65  (0.72)} &  \textbf{0.62  (0.69)} & \textbf{0.41  (0.46)} & \textbf{0.41  (0.44)}\\
120 $h^{-1}$Mpc & \textbf{0.48  (0.52)} &  \textbf{0.51  (0.54)} & \textbf{0.38  (0.39)} & \textbf{0.39  (0.39)}\\
\cline{2-5}
& \multicolumn{4}{c}{Peak histogram reduced $\chi^2$}\\
\cline{2-5}
20 $h^{-1}$Mpc  & 14.50 (17.12) & 11.33 (13.52) &  5.74  (6.34) & 4.78  (5.40)\\
40 $h^{-1}$Mpc  & 2.08  (2.41) &  1.87  (2.17) &  \textbf{0.97}  (1.02) & \textbf{0.89  (0.95)}\\
60 $h^{-1}$Mpc  &  \textbf{0.61  (0.62)} &  \textbf{0.61  (0.60)} & \textbf{0.37  (0.33)} & \textbf{0.36  (0.32)}\\
120 $h^{-1}$Mpc &  \textbf{0.56  (0.53)} &  \textbf{0.56  (0.52)} & \textbf{0.36  (0.37)} & \textbf{0.36  (0.37)}\\
\cline{2-5}
& \multicolumn{4}{c}{$V_0$ reduced $\chi^2$}\\
\cline{2-5}
20 $h^{-1}$Mpc  & 58.55 (82.06) & 32.13 (44.96) &  18.16  (25.01) & 13.17  (18.27)\\
40 $h^{-1}$Mpc  &  5.38  (7.39) &  4.37  (6.01) &  2.16  (2.79) & 1.95  (2.54)\\
60 $h^{-1}$Mpc  &  \textbf{0.99}  (1.19) &  \textbf{0.97}  (1.16) & \textbf{0.54  (0.58)} & \textbf{0.54  (0.57)}\\
120 $h^{-1}$Mpc &  \textbf{0.70  (0.81)} &  \textbf{0.81  (0.94)} & \textbf{0.39  (0.41)} &  \textbf{0.40  (0.41)}\\
\cline{2-5}
& \multicolumn{4}{c}{$V_1$ reduced $\chi^2$}\\
\cline{2-5}
20 $h^{-1}$Mpc  & 56.34 (78.72) & 41.79 (58.42) &  19.70  (26.83) & 13.89  (19.10)\\
40 $h^{-1}$Mpc  & 6.61  (9.04) &  5.78  (7.93) &  2.25  (2.93) &  2.05  (2.68)\\
60 $h^{-1}$Mpc  & 1.20  (1.47) &  1.16  (1.42) &  \textbf{0.60  (0.56)} & \textbf{0.62  (0.57)}\\
120 $h^{-1}$Mpc & \textbf{1.87  (2.40)} &  \textbf{2.78  (3.66)} & \textbf{0.41  (0.38)} &  \textbf{0.41  (0.37)}\\
\cline{2-5}
& \multicolumn{4}{c}{$V_2$ reduced $\chi^2$}\\
\cline{2-5}
20 $h^{-1}$Mpc  & 46.94 (64.68) & 33.04 (45.63) & 18.26  (24.98) & 13.32  (18.30)\\
40 $h^{-1}$Mpc  &  5.26  (7.00) &  4.53  (6.05) &  2.13  (2.79) & 1.96  (2.57)\\
60 $h^{-1}$Mpc  &  1.11  (1.25) &  1.11  (1.23) &  0.61  (0.45) & \textbf{0.62  (0.44)}\\
120 $h^{-1}$Mpc &  1.43  (1.61) &  1.67  (1.94) &  \textbf{0.40  (0.32)} &  \textbf{0.40  (0.31)}\\
\hline
\hline
\end{tabular}
\caption{Goodness of fit for different lens plane thickness configurations, based on the reduced $\chi^2$ (i.e. $\chi^2$ per degree of freedom).  Configurations that yield good fits ($\chi^2 \leq 1$), implying that they are indistinguishable from the fiducial case, are highlighted in bold. \textbf{Power spectrum:} values for a range of $\ell \in [200, 12000]$ and in parenthesis $\ell \in [200, 3532]$. \textbf{PDF:} values for a range of $\kappa \in [-3.0, 5.0]$ in units of the shape noise r.m.s., and in parenthesis $\kappa \in [-3.0, 3.1]$. \textbf{Peak counts:} values for a range of $\kappa \in [-0.5, 5.0]$ in units of the shape noise r.m.s., and in parenthesis $\kappa \in [-0.5, 4.0]$. \textbf{MFs:} values for a range of $\kappa \in [-2.0, 5.0]$ in units of the shape noise r.m.s., and in parenthesis $\kappa \in [-2.0, 3.0]$.}
\label{table:chi2_planes}
\end{center}
\end{table*}

\subsection{Mass resolution} \label{subsecsec:resolution}

\begin{figure*}
\begin{center}
\includegraphics[width=\linewidth]{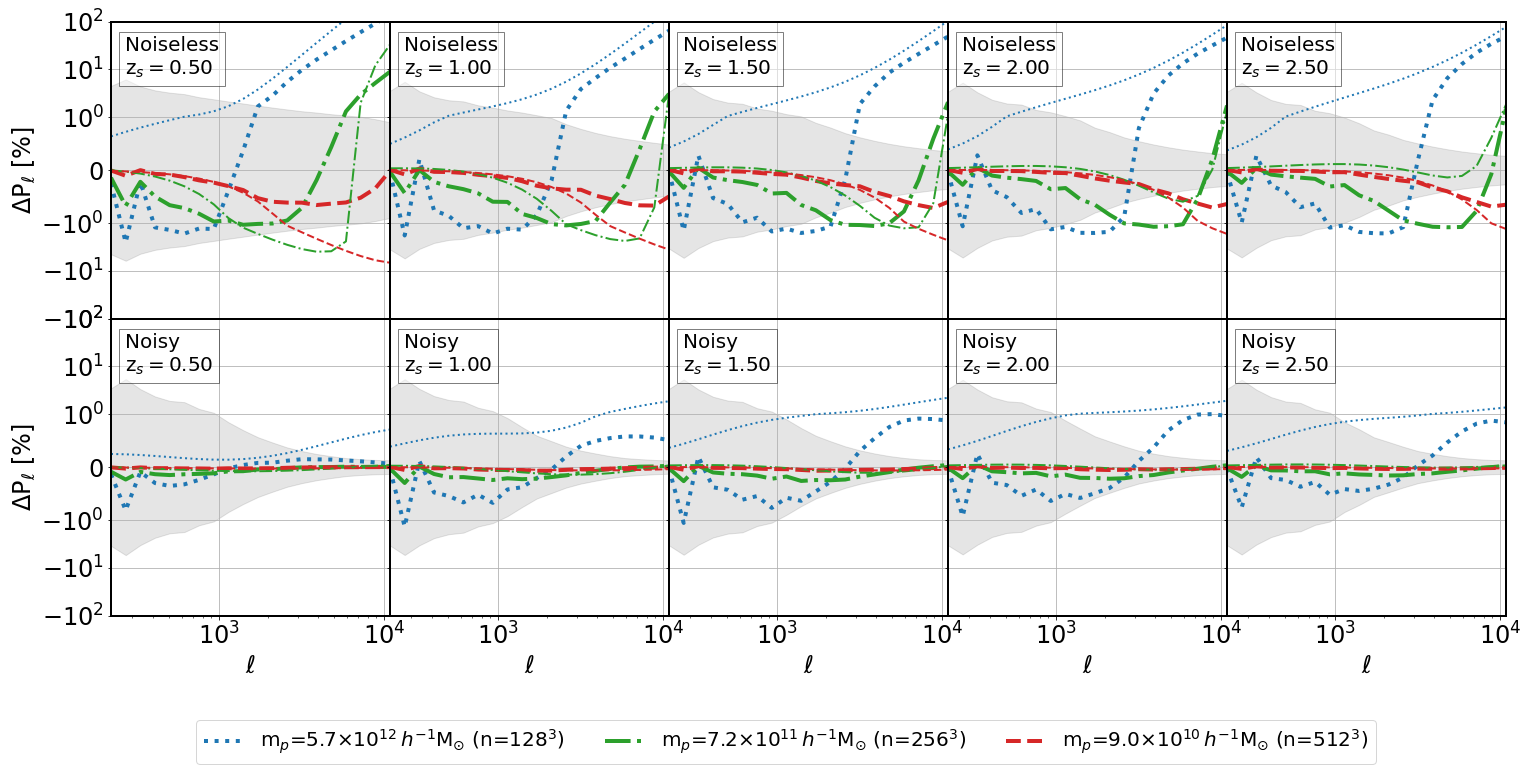}
\caption{Effect of mass resolution (number of particles in the simulation volume) on the convergence power spectrum. Each column shows, for a different redshift bin, the percentage difference in the mean auto power spectrum measured over 10,048 convergence maps for a given mass resolution of the underlying N-body simulation, relative to the fiducial value of $1.1\times10^{10}\,{\rm M_{\odot}}$ (corresponding to $1024^3$ DM particles in the simulation box). The top row corresponds to noiseless data, and the bottom row to data in the presence of shape noise (in all cases, maps were smoothed at $1\,{\rm arcmin}$ resolution). Thick lines represent measurements over simulated maps, while thin lines represent theoretical predictions following Eq.~\eqref{eq:ps_particles}. For comparison, a standard error is shown in shaded gray, corresponding to 3 standard deviations of the measurements in the fiducial case, scaled to a $2\times10^4$\,deg$^2$ survey.
}
\label{fig:particlemass_ps}
\end{center}
\end{figure*}

Figure~\ref{fig:particlemass_ps} displays the mean percentage difference in the power spectrum for noiseless and noisy maps for different mass resolutions, compared to the highest-resolution case. The number of particles in the four configurations we tested are $128^3$, $256^3$, $512^3$ and $1024^3$. As done  in Figure~\ref{fig:planethickness_ps} with the sensitivity to the lens plane thickness, a standard error corresponding to 3 standard deviations for the fiducial case is shaded for reference in Figure~\ref{fig:particlemass_ps}.

The main difference between the power spectra is an increase in power on small scales, with a relative loss of power on intermediate scales, as the mass resolution decreases. Qualitatively, this behavior matches what is expected from the matter power spectrum: a loss of power at intermediate scales as the mass resolution decreases (captured by the first term in Eq.~\eqref{eq:ps_particles}) followed by a sharp increase at small scales due to shot noise. Similar effects have been described in past studies \citep{Jain2000, Vale2003}. We cannot fit all the data with the simple model in Eq.~\eqref{eq:ps_particles}, though. For example, a smoothing scale of $\sigma_N=5\%$ of the average inter-particle separation gives results that are qualitatively in line with the relative power change when going from $256^3$ to $1024^3$ particles, but not for the other resolutions tested. An effect not taken into consideration in that simple model is that of the softening length on the relaxation time of the simulations. All of our simulations kept that length constant, and the softening for the lower resolution runs may have been insufficient. Another effect that can induce a loss in power is the reduction in the linear growth factor due to the discreteness of the initial conditions, as described in ~\citep{Heitmann2010}. The presence of shape noise, which dominates at small scales, mitigates both effects, rendering the measured statistic more forgiving to the resolution of the underlying N-body simulation.

\begin{figure*}
\begin{center}
\includegraphics[width=\linewidth]{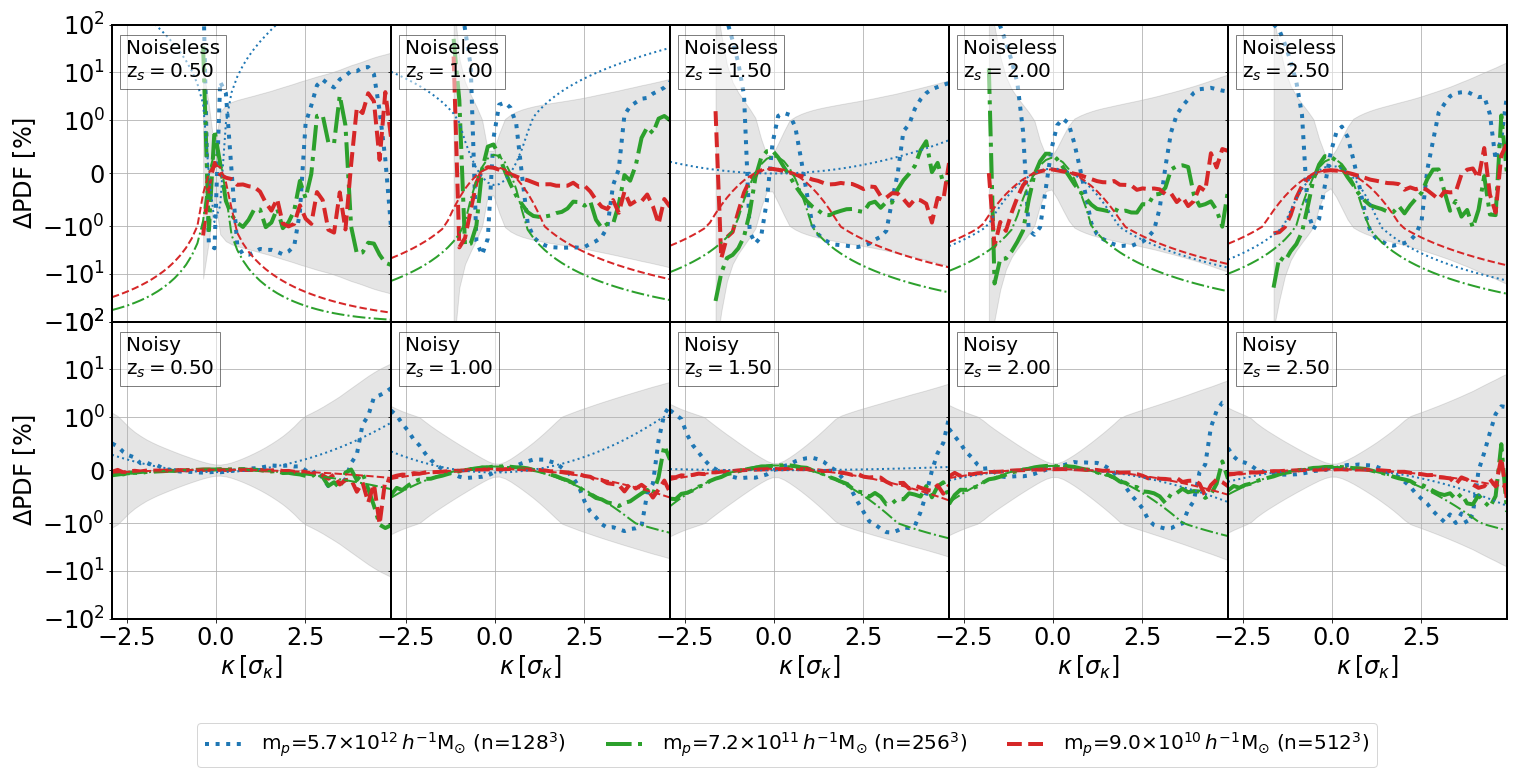}
\caption{Same as Figure~\ref{fig:particlemass_ps} for the one-point $\kappa$ probability density function (PDF). Thick lines represent measurements over simulated maps, while thin lines represent the expectation for GRFs with the same power spectrum as the $\kappa$ maps.
}
\label{fig:particlemass_pdf}
\end{center}
\end{figure*}

The effect of mass resolution on the one-point probability density function is not as straightforward as for the lens plane thickness. The GRF model does not work particularly well for noiseless data, as can be seen in Figure~\ref{fig:particlemass_pdf}. Except for the lowest-resolution simulation, shape noise still keeps any differences within the standard error of the simulations themselves. In the absence of noise, $512^3$ and $1024^3$ particles seem to yield equivalent results.

\begin{figure*}
\begin{center}
\includegraphics[width=\linewidth]{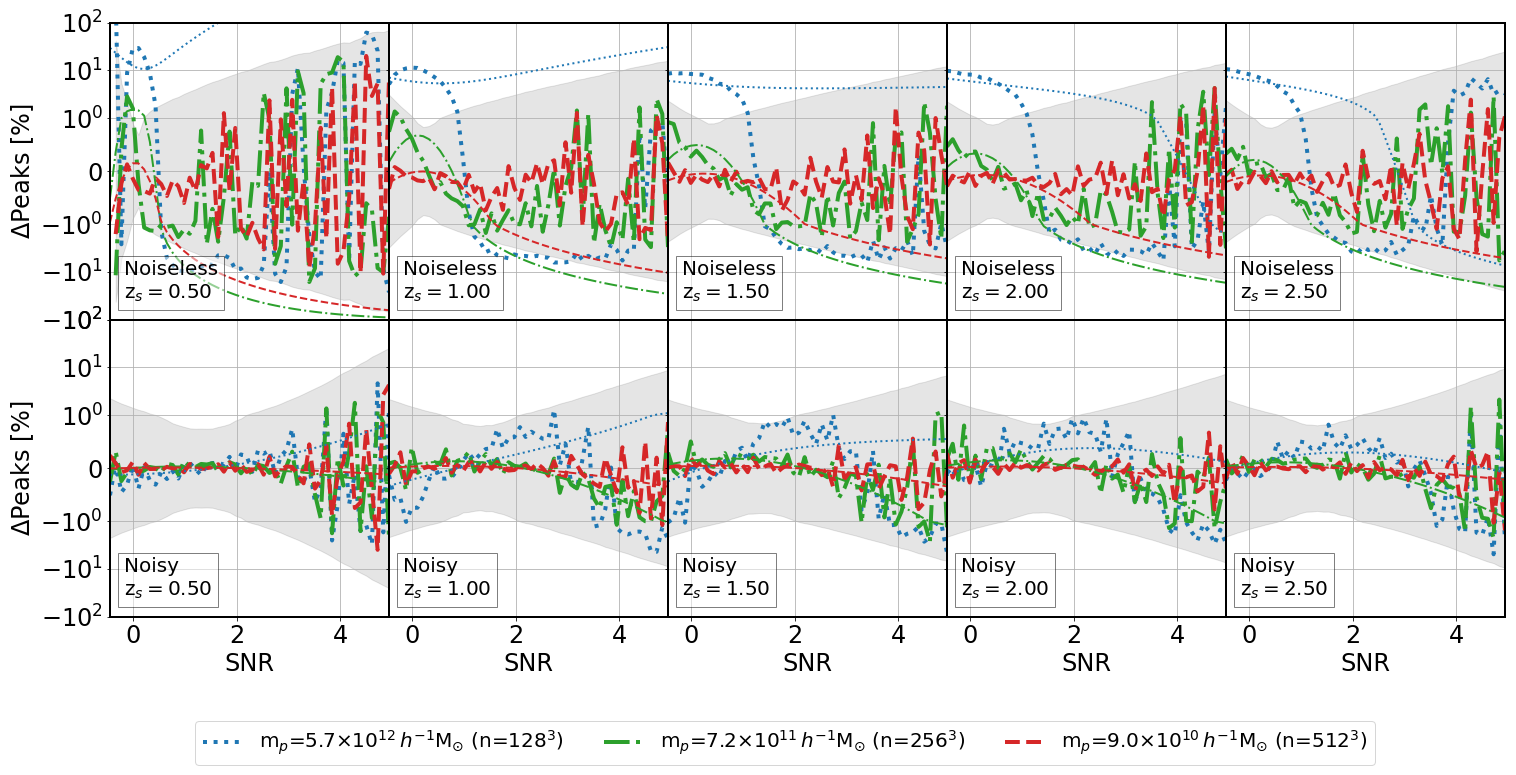}
\caption{Same as Figure~\ref{fig:planethickness_ps} for the lensing peak distribution. Thick lines represent measurements over simulated maps, while thin lines represent the expectation for GRFs with the same power spectrum as the $\kappa$ maps.
}
\label{fig:particlemass_peaks}
\end{center}
\end{figure*}

Lensing peaks are a more robust statistic to the mass resolution (see Figure~\ref{fig:particlemass_peaks}), and the configurations with $256^3$ and $512^3$ particles are statistically indistinguishable from the $1024^3$ fiducial for both noisy and noiseless data. The model with the lowest mass resolution yields histograms whose differences from the fiducial case clearly exceed the statistical uncertainty. The difference is more important for low-significance peaks, which can be induced by the Poisson shot noise present in the underlying N-body simulation. Shape noise reduces the differences at the low significance tail, where peaks are noise-dominated. The GRF model for peak counts seems to work better than for the PDF.

\begin{figure*}
\begin{center}
\includegraphics[width=\linewidth]{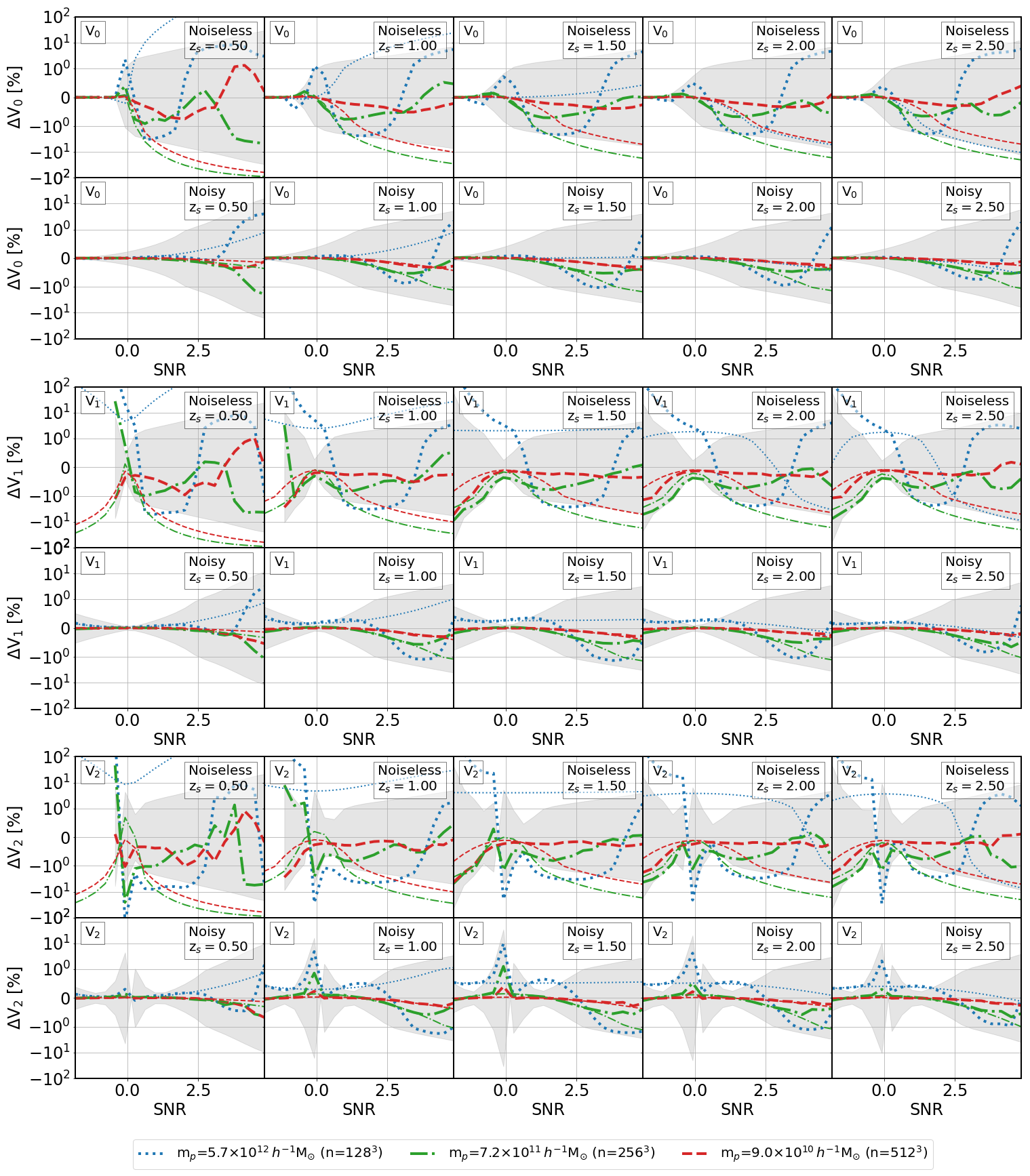}
\caption{Same as Figure~\ref{fig:planethickness_ps} for the three Minkowski functionals (MFs). Each pair of rows shows a different MF. The top row of each pair corresponds to noiseless data, and the bottom row to data in the presence of shape noise (in all cases, maps were smoothed at $1\,{\rm arcmin}$ resolution). Thick lines correspond to measurements over simulated data, thin lines to predictions for Gaussian random fields with the same power spectrum as the simulated maps.
}
\label{fig:particlemass_mfs}
\end{center}
\end{figure*}

The effect of mass resolution on the measured MFs is diplayed in Figure~\ref{fig:particlemass_mfs}. As with the PDF, a GRF model does not fit as well the impact of changes in mass resolution as it did for changes in lens plane thickness. This may be partly due to the relatively larger change in power, as well as the $\ell$-dependence of that change. In general, measurements performed on simulations are less sensitive than what would be expected from GRFs, and their changes remain within errors for most cases.

Table~\ref{table:chi2_particlenumber} collects the statistical significance of changes in the mass resolution of the N-body simulations used to build synthetic $\kappa$ maps on the observables under study. Most of the considerations discussed in \S~\ref{subsec:thickness} about differences in the covariance matrices for the observables apply here as well. For example, the differences for the power spectrum covariance are modest with differences in the dominant, diagonal terms, of less than 3\% between the covariances for the $256^3$ and $1024^3$ particle simulations. For analyses in the presence of shape noise levels commensurate with the ones considered in this study, $256^3$ particles suffice. If the multipole range is limited to $\approx 3500$, even lower resolutions can be used for the power spectrum (or alternatively the same $256^3$ resolution can be used for negligible levels of shape noise). The non-Gaussian statistics that are the most sensitive to resolution effects are the MFs.

As with the results for the thickness of the lensing planes, the figures in Table~\ref{table:chi2_particlenumber} do not incorporate the possible effect of errors in the covariance matrices used in the calculation of $\chi^2$. In this case, the Minkowski functionals are also the most sensitive statistic. For their measurements on the $256^3$ particles' simulation to be statistically different than the measurements on the $1024^3$ particles' maps, the true covariance would need to be overestimated by more than $117\%$.

\begin{table*}
\begin{center}
\begin{tabular}{lcccc}
\hline
\hline
& \multicolumn{2}{c}{Noiseless} & \multicolumn{2}{c}{Noisy}\\
\cline{2-5}
\multicolumn{1}{c}{Particle mass} & \multicolumn{1}{c}{Model-dependent} & \multicolumn{1}{c}{Fiducial} & \multicolumn{1}{c}{Model-dependent} & \multicolumn{1}{c}{Fiducial}\\
\multicolumn{1}{l}{(number or particles)} & & & & \\ 
\hline
& \multicolumn{4}{c}{Power spectrum reduced $\chi^2$}\\
\cline{2-5}
$5.7\times10^{12}~h^{-1}$ $M_{\odot}$ $(128^3)$ & 11494.43 (25.03) & 44822.28 (25.69) & 42.44 \textbf{(0.85)} & 44.26 \textbf{(0.85)}\\
$7.2\times10^{11}~h^{-1}$ $M_{\odot}$ $(256^3)$ & 48.97 \textbf{(0.63)} & 50.48 \textbf{(0.62)} & \textbf{0.17 (0.10)} & \textbf{0.17 (0.10)} \\
$9.0\times10^{10}~h^{-1}$ $M_{\odot}$ $(512^3)$ & 1.68 \textbf{(0.05)} & 1.66 \textbf{(0.04)} & \textbf{0.03 (0.01)} & \textbf{0.03 (0.01)}\\
\cline{2-5}
& \multicolumn{4}{c}{PDF reduced $\chi^2$}\\
\cline{2-5}
$5.7\times10^{12}~h^{-1}$ $M_{\odot}$ $(128^3)$  & 67.96 (89.38)  &  289.44 (382.24)  &  \textbf{0.72 (0.85)}  &  \textbf{0.71 (0.83)}\\
$7.2\times10^{11}~h^{-1}$ $M_{\odot}$ $(256^3)$  &  1.33   (1.66) & \textbf{0.16   (0.16)} & \textbf{0.16  (0.16)} & \textbf{0.16  (0.16)}\\
$9.0\times10^{10}~h^{-1}$ $M_{\odot}$ $(512^3)$  &  \textbf{0.10   (0.06)} & \textbf{0.10   (0.06)} & \textbf{0.05  (0.03)} & \textbf{0.05  (0.04)}\\
\cline{2-5}
& \multicolumn{4}{c}{Peak histogram reduced $\chi^2$}\\
\cline{2-5}
$5.7\times10^{12}~h^{-1}$ $M_{\odot}$ $(128^3)$  & 404.38 (488.55)  &  426.60 (513.32)  &  1.44 (1.68)  &  1.44 (1.68)\\
$7.2\times10^{11}~h^{-1}$ $M_{\odot}$ $(256^3)$  &  \textbf{0.74   (0.84)} & \textbf{0.76   (0.86)} & \textbf{0.24  (0.24)} & \textbf{0.24  (0.24)}\\
$9.0\times10^{10}~h^{-1}$ $M_{\odot}$ $(512^3)$  &  \textbf{0.29   (0.31)} & \textbf{0.28   (0.30)} & \textbf{0.07  (0.07)} & \textbf{0.08  (0.07)}\\
\cline{2-5}
& \multicolumn{4}{c}{$V_0$ reduced $\chi^2$}\\
\cline{2-5}
$5.7\times10^{12}~h^{-1}$ $M_{\odot}$ $(128^3)$  & 53.02 (70.59)  & 89.27 (119.06)  & 1.47 (1.71)  &  1.47 (1.70)\\
$7.2\times10^{11}~h^{-1}$ $M_{\odot}$ $(256^3)$  &   \textbf{0.68   (0.76)} & \textbf{0.69   (0.77)} & \textbf{0.16  (0.19)} & \textbf{0.16  (0.20)}\\
$9.0\times10^{10}~h^{-1}$ $M_{\odot}$ $(512^3)$  &  \textbf{0.07   (0.05)} & \textbf{0.07   (0.05)} & \textbf{0.02  (0.02)} & \textbf{0.02  (0.02)}\\
\cline{2-5}
& \multicolumn{4}{c}{$V_1$ reduced $\chi^2$}\\
\cline{2-5}
$5.7\times10^{12}~h^{-1}$ $M_{\odot}$ $(128^3)$  & 1273.03 (1691.45)  &  2103.48 (2796.06)  &  8.83 (11.26)  & 8.81 (11.29)\\
$7.2\times10^{11}~h^{-1}$ $M_{\odot}$ $(256^3)$  &  15.19   (20.21) & 16.15   (21.47) & \textbf{0.46  (0.58)} & \textbf{0.46  (0.58)}\\
$9.0\times10^{10}~h^{-1}$ $M_{\odot}$ $(512^3)$  &  \textbf{0.70   (0.89)} & \textbf{0.69   (0.88)} & \textbf{0.07  (0.08)} & \textbf{0.07  (0.08)}\\
\cline{2-5}
& \multicolumn{4}{c}{$V_2$ reduced $\chi^2$}\\
\cline{2-5}
$5.7\times10^{12}~h^{-1}$ $M_{\odot}$ $(128^3)$  & 5884.23 (7837.70)  &  13038.65 (17416.22)  &  4.37 (5.46)  &  4.40 (5.51)\\
$7.2\times10^{11}~h^{-1}$ $M_{\odot}$ $(256^3)$  &  126.88   (168.00) & 137.14   (182.01) & \textbf{0.29  (0.34)} & \textbf{0.29  (0.35)}\\
$9.0\times10^{10}~h^{-1}$ $M_{\odot}$ $(512^3)$  &  \textbf{0.84}   (1.08) & \textbf{0.83}   (1.07) & \textbf{0.04  (0.04)} & \textbf{0.04  (0.04)}\\
\hline
\hline
\end{tabular}
\caption{Goodness of fit as in Table~\ref{table:chi2_planes}, but for different mass resolution configurations. Configurations with $\chi^2 \leq 1$, implying that they are indistinguishable from the fiducial case, are highlighted in bold. \textbf{Power spectrum:} values for a range of $\ell \in [200, 12000]$ and in parenthesis $\ell \in [200, 3532]$. \textbf{PDF:} values for a range of $\kappa \in [-3.0, 5.0]$ in units of the shape noise r.m.s., and in parenthesis $\kappa \in [-3.0, 3.1]$. \textbf{Peak counts:} values for a range of $\kappa \in [-0.5, 5.0]$ in units of the shape noise r.m.s., and in parenthesis $\kappa \in [-0.5, 4.0]$. \textbf{MFs:} values for a range of $\kappa \in [-2.0, 5.0]$ in units of the shape noise r.m.s., and in parenthesis $\kappa \in [-2.0, 3.0]$.}
\label{table:chi2_particlenumber}
\end{center}
\end{table*}

\section{Conclusions} \label{sec:conclusions}
We performed a series of numerical experiments to test the influence of the lens plane thickness and the mass resolution of ray-traced N-body simulations on commonly used WL statistics: the convergence power spectrum, the one-point probability density function, lensing peak counts, and Minkowski functionals. While our simulations cannot be used directly to analyze survey data, they can serve to guide design choices in studies of non-Gaussian statistics on small scales. They set some minimal requirements to avoid significant biases in predictions obtained from the application of the multiple lens plane algorithm. 

We have found that in the multi-plane ray-tracing algorithm, lens planes in the range $60-120~h^{-1}$\,Mpc are safe to use in analysis of WL data with galaxy densities commensurate with those in LSST-like surveys. Thinner planes will result in a loss of power across a wide range of multipoles, that can in principle be accounted for analytically for the power spectrum but not for non-Gaussian statistics. On the thick-plane end, there are no biases induced in the observables, and the computational time for raytracing past light cones can be reduced (e.g. by $\approx 33\%$ when $120~h^{-1}$\,Mpc planes are used instead of $80~h^{-1}$\,Mpc). However, such a choice implies a larger number of N-body simulations to generate the same number of pseudo-independent realizations of the $\kappa$ field, increasing the overall computation budget.

In order to analyze WL data sets at angular resolutions of 1\,arcmin with LSST levels of shape noise, simulations with mass resolutions of $7.2\times 10^{11}$\,$M_{\odot}$ per DM particle are sufficient (corresponding to $256^3$ particles in a $240~h^{-1}$\,Mpc simulation box), even if non-Gaussian statistics are included in the analysis. Moving, for instance from $512^3$ to $256^3$ particles could bring computational time savings of $\approx 77\%$.

\acknowledgments
\section*{Acknowledgments}
We thank the anonymous referee for providing insightful feedback that has resulted in an improved manuscript. We also thank Alexandre Refregier for useful comments on the original manuscript and Takashi Hamana for pointing us to an analytic model for the loss of power due to finite-thickness lens planes. The simulations and data analysis for this work were performed on NSF's XSEDE facility. This work was also supported by NASA ATP grant 80NSSC18K1093. 

\bibliography{refs}

\end{document}